\documentclass[letterpaper,10pt,conference,dvipsnames]{ieeeconf} 
\let\labelindent\relax
\IEEEoverridecommandlockouts                          
\usepackage{cite}
\usepackage{enumitem}
\usepackage{amsmath,amssymb,amsfonts}
\usepackage{algorithmic}
\usepackage{graphicx}
\usepackage{textcomp}
\usepackage{subfigure}
\usepackage{xcolor}
\usepackage{theorem}
\usepackage[font=footnotesize,skip=0.1pt,justification=centering]{caption}
\usepackage{tikz}
\usetikzlibrary{shapes.geometric,decorations.pathreplacing}
\usepackage{pgfplots,pgfplotstable}
\usepgfplotslibrary{fillbetween}
\pgfplotsset{compat=1.4}
\usetikzlibrary{pgfplots.groupplots}
\usepackage{pgf}
\usepackage[draft]{todonotes}
\usepackage{enumitem}
\usepackage{booktabs}
\usepackage{url}
\definecolor{MyRed}{HTML}{e6550d}
\definecolor{EPFLRed}{HTML}{b51f1f}
\definecolor{MyBlue}{rgb}{0.20, 0.6, 0.78}
\definecolor{MyGreen}{rgb}{0.4,0.8,0.4}

\DeclareMathOperator{\Z}{\mathbb{Z}}
\DeclareMathOperator{\I}{\mathcal{I}}

\newcommand{\change}[1]{\textcolor{black}{#1}}

\newtheorem{remark}{Remark}

\allowdisplaybreaks[4]

\begin{document}

\title{\LARGE \bf Scheduling Delays and Curtailment for Household Appliances with Deterministic Load Profiles using MPC}

\author{Yingzhao Lian, Yuning Jiang$^*$, Colin N. Jones, and Daniel F. Opila 
\thanks{This work was support from the Swiss National Science Foundation under the RISK project (Risk Aware Data-Driven Demand Response), grant number 200021 175627. (Corresponding author: Yuning Jiang)}
\thanks{YL, YJ and CJ are with Automatic Control Laboratory, EPFL. (email: \tt{yingzhao.lian, yuning.jiang,colin.jones}@epfl.ch)}
\thanks{DO is with United States Naval Academy, Electrical and Computer Engineering Department. (email: \tt{opila}@usna.edu)}
}

\maketitle
\begin{abstract}
Smart home appliances can time-shift and curtail their power demand to assist demand side management or allow operation with limited power, as in an off-grid application. \change{This paper proposes a scheduling process to start appliances with time-varying deterministic load profiles. Self-triggered model predictive control is used to limit the household net power demand below a given threshold.} Meanwhile, deterministic load profiles are more difficult to schedule compared to variable charging or thermal loads \change{because system failure will occur once power demand is not satisfied}. The proposed scheme formulates the decision of the load shifting time as a continuous optimization problem, and an inhomogeneous time grid system is introduced to handle the optimization of different appliances and their consensus at this resolution. The efficacy of the proposed scheme is studied by numerical comparison with a mixed-integer MPC controller and by a case study of three home appliances and an interruptible washing machine.
\end{abstract}
\change{
}

\vspace{-.3em}
\section{Introduction}
The residential sector is a large energy consumer, accounting for~28\% of the total energy consumption in Europe, and scheduling this load plays a key role in demand side management (DSM)~\cite{vazquez2010energy}. \change{DSM is an arrangement of actions to modify the power demand on the user side, and load scheduling for home appliances has experimentally shown a peak load reduction up to 40\%~\cite{erol2011wireless}.} Thus, this method can improve conditions in areas suffering from power shortages, such as Pakistan~\cite{baloch2016current} and Lebanon~\cite{fardoun2012electricity}. Load scheduling of intelligent home appliances is popular even aside from power considerations, based on an online survey conducted in Europe~\cite{stamminger2017load}. In particular, 50\% of the participants preferred to use the scheduling function, 29\% of them mentioned a reduced electricity tariff, and 7\% got a stability improvement in home power management. 

\change{Motivated by these benefits, the load scheduling problem has been widely studied, and model predictive control (MPC) is an attractive solution because it can enforce explicit constraint satisfaction by synthesizing the system behavior forecast recursively~\cite{rakovic2018handbook}. The control performance is varied by the design of the objective, which can reflect the occupants' comfort and operating time preference~\cite{godina2018model,mariano2021review}.}


\change{In this work, we consider loads whose demand follows a specific time-varying profile once started, termed a ``deterministic'' power demand profile. Unlike the continually variable loads often considered in existing work, such as vehicle charging~\cite{hou2019smart} or an averaged on-off thermal loads~\cite{lian2021adaptive}, the loads considered here can be delayed by a time shift, curtailed, or interrupted at specific break points, but their demand follows the profile when running. Among these actions, time shifts pose the major challenge, which makes the time of different loads not aligned and therefore nonlinear.} To address this problem, most previous results formulate the load profile of an appliance as a discrete-time sequence, where the time shift is an integer-valued decision variable~\cite{9147361,di2012model,yang2014joint,qayyum2015appliance}. Moreover, due to the computational complexity of this mixed-integer problem, most previous work does not consider constraints and load curtailment except for~\cite{9147361}.



\change{This paper proposes a new continuous formulation, which is inspired by self-triggered MPC.} Self-triggered MPC can systematically optimize the active/triggered time of a device proactively with explicit constraint enforcement~\cite{heemels2012introduction}, and has been widely used in sensor networks and Internet-of-Things~\cite{stricker2021distributed}. \change{As opposed to the mixed-integer formulation~\cite{berglind2012self,bernardini2012energy}, the triggering time can be a continuous decision variable~\cite{lian2020resource,lian2021resource,lian2022resource} in this scheme}. This paper interprets the time shift as a trigger time viewed from the current time step, and transcribes the load scheduling problem into a continuous optimization problem such that we can take full advantage of continuous numerical solvers.


The contributions are summarized as follows: (a) A continuous formulation of the load scheduling problem with time shifts. This framework is compatible with interruptible loads with explicit break points, and its computational complexity caused by the shifting decision is constant; (b) An inhomogeneous time grid system is introduced to enable control of the computational complexity of each load and constraint; (c) The system demand respects a time-varying power limit, \change{as would occur off-grid or in microgrids, or if the utility controls demand via a power limit signal rather than a price.}
    
    
    

The paper is organized as follows: the load scheduling problem and its corresponding optimization problem are introduced in Section~\ref{sect:model}. The numerical implementation into a continuous tractable formulation is introduced in Section~\ref{sect:num}, alongside a comparison with previous mixed integer formulations. The numerical results and a comparison with the mixed-integer formulation are in Section~\ref{sect:result}, followed by conclusions in Section~\ref{sect:conclusion}.

\textit{Notations}: $\Z_{a}^b$ denotes the set of integers ranging from $a$ to $b$ with $\Z$ the set of non-negative integers. $\{x_i\}_{i\in T}$ denotes a set indexed by set $T$, and the index is dropped when there is no confusion as $\{x_i\}$. $\mathcal A\backslash \mathcal B:=\{x|x\in\mathcal A,\;x\notin \mathcal B\}$.

\section{Problem Statement}\label{sect:model}

Here we consider the coordination of multiple deterministic loads, appliances for example. This is relevant for varying electricity prices, but is a particular problem when both power and energy are limited as in off-grid operation, a low-power connection, or when responding to a demand limit signal from a grid operator. We consider the case of a dwelling with a battery and a limited external power source like solar, fuel cell, etc. \change{ which can reduce its output if needed. Additionally, the primary goal is to ensure feasibility and to minimize delays, in contrast to a typical objective of minimizing energy costs with no limit on power.}

Once an appliance is requested to start, it should launch as soon as possible considering the feasible power profile, the total power limit, and total available energy. In the rest of this paper, we use absolute time in all notation.

\subsection{The Load Profile Models}
Two different load types are considered: uninterruptible loads, whose execution cannot be paused once started, and interruptible loads with explicit break points. An example for the latter case is a washing machine with a drying function, which can pause after rinsing with a reasonable delay. If we consider an uninterruptible load, indexed by $i$, launched at time $t_{s_i}$, its actual power consumption $\overline{u}_i(t)$ is expected to follow a predefined nominal power profile defined by
\begin{align}\label{eqn:load_model}
   \begin{cases}
    \overline{u}_i(t),\; &t\in [t_{s_i},t_{s_i}+T_{e_i}],\\
    0,\; & \text{otherwise},
\end{cases} 
\end{align}
where $T_{e_i}$ is the time span of the execution of $i$-th load.

Similarly, an interruptible load with explicit break points can be modelled as a sequence of uninterruptible subtasks. In particular, we define $\mathcal{I}_s$, the set that includes all the uninterruptible loads and the first subtask of the interruptible loads. This means that if an index $j$ represents a subtask of an interruptible load but not the first, then we have $j\notin\mathcal{I}_s$, and accordingly we can define a mapping $\mathfrak{p}(j)$ that returns the finishing time of its precedent subtask for all $j\notin\I_s$. Taking a washing machine with a drying function as an example, we can model it as two uninterruptible subtasks indexed by $i$ and $j$, then we have $i\in\mathcal{I}_s$ and $j\notin\mathcal{I}_s$ such that $\mathfrak{p}(j)=t_{s_i}+T_{e_i}$. Since each subtask must start after the earlier one finishes, $t_{s,j}\geq\mathfrak{p}(j)$ always holds for all $j\notin\mathcal{I}_s$.



\subsection{The Scheduling Strategies}
The external power source nominally provides two functions, power to support the instantaneous loads and energy over time to prevent battery discharge.  Distributed generation sources are variable and difficult to predict, and we can consider $P_{\min}(t)$ as the minimum power that can be reliably expected at a given time, and $P_\mathrm{avg}$ as the expected average power over an interval $[t_0,t_0+T]$. While it is often true that  $P_\mathrm{avg}>P_\mathrm{min}(t)$, the uncertainty limits reliable operation, hence we only consider $P_\mathrm{min}(t)$ in this study. The excess power from the external source will be stored in the battery and allow for more power use later without resorting to the grid. 


Multiple appliances can run concurrently, and, due to the discussion about $P_\mathrm{min}(t)$ and the battery above, the total power consumption from these active appliances are required to be upper bounded by $u_\mathrm{max}(t)$ as
\begin{align}\label{eqn:bus_limit}
    \sum_{i\in\I(t)} u_i(t) \leq u_\mathrm{max}(t) := P_\mathrm{batt}(t)+P_\mathrm{min}(t),
\end{align}
where $\I(t)$ is the set of all active appliances at time $t$ and $P_\mathrm{batt}$ denotes the power supplied by the battery, whose charge level is modelled by
\begin{align*}
    \frac{dE_\mathrm{batt}(t)}{dt} = -P_\mathrm{batt}(t)\;.
\end{align*}
The battery capacity is bounded within $[0,E_\mathrm{max}]$. Its charging/discharging rate is limited such that $P_\mathrm{batt}(t)\in[\underline{P}_\mathrm{batt},\overline{P}_\mathrm{batt}]$ with $\overline{P}_\mathrm{batt}/\underline{P}_\mathrm{batt}$ the maximal dis-/charge rate. It is possible to consider the charging/discharging efficiency based on the method given in~\cite{di2012model}, but for the sake of clarity we consider an ideal battery. \change{Additionally, the capacity and allowable dis-/charge rate of the battery can affect the performance of the system~\cite{sinsley2020upper}.}

To start the appliances as early as possible after their requests without violating this constraint, shifting and curtailment are considered:
\subsubsection{Load shifting}
The execution of multiple loads can be shifted after their request to avoid overlap of the peak power consumption. If an appliance is requested at $t_{r_i}$, the delay between the request and the actual starting time of the appliance is required to be bounded by $\Delta_i^d$,
\[
0\leq t_{s_i}-t_{r_i}\leq \Delta_{i}^d.
\]
The preference of a responsive execution is modelled by a monotonically increasing loss function $J_{i}^d(t_{s_i}-t_{r_i})$.
\subsubsection{Load curtailment} The load applied $u_i(t)$ is allowed to mismatch the nominal load profile $\overline{u}_i(t)$. The resulting curtailment is required to be bounded as
\begin{align*}
    \forall\,t\in[t_{s_i},t_{s_i}+T_{e_i}],\;0\leq \overline{u}_i(t) - u_i(t)\leq c_{i}(t),
\end{align*}
where $c_i(t)$ is the maximal curtailment at time $t$. Accordingly, the preference of matching the nominal profile is modelled by objective $J_{i}^c(\cdot)$, the corresponding accumulated curtailment cost is given by
\begin{equation}\label{eq::L}
    L_{i}^c(u_i(\cdot),t_{s_i}):=\int_{t_{s_i}}^{t_{s_i}+T_{e_i}}J_{i}^c(\overline{u}_i(t)-u_i(t))\mathrm{d}t.
\end{equation}

Based on the aforementioned scheduling strategies, at time instance $t$, the home appliances are characterized by two sets, the set of appliances that have been requested to launch, dubbed $\mathcal{I}_r(t)$ and the set of appliances that are executing currently at time $t$, dubbed $\mathcal{I}(t)$. In particular, $\mathcal{I}_r(t)$ includes all the loads that have been requested to launch and are not yet finished i.e., ($t\geq t_{r_i}$), as well as its subsequent loads if it is part of an interruptible load. Note that a requested load can be delayed, thus we also have $\mathcal{I}(t)\subset\mathcal{I}_r(t)$.
\begin{remark}
Note that the proposed scheme can also handle curtailment cost \change{$J_{i}^c$} as a function of time. This scenario appears when portions of the profile are more critical. \change{Meanwhile, the priority of a specific load can be reflected in the penalty in its corresponding $J_{i}^c$ and $J_i^d$.}
\end{remark}

\subsection{Load Scheduling MPC}
At $t_0$, given the list of requested loads $\mathcal{I}_r(t_0)$, we can summarize the load scheduling MPC problem as:
\begin{subequations}\label{eqn:opt_inf}
\begin{align}
    \hspace{-2mm}\underset{t_{s_i},P_\mathrm{batt}(\cdot),u_i(\cdot)}{\mathrm{minimize}} & \sum_{i\in\I_r(t_0)} J_{i}^d(t_{s_i}-t_{r_i})+L_i^c(u_i(\cdot),t_{s_i})
\end{align}
subject to
\begin{align}
    &\sum_{i\in\I_r(t_0)} u_i(t) \leq P_\mathrm{min}(t)+P_\mathrm{batt}(t),\;t\in [t_0,t_e],\label{eqn:opt_bus_cons}\\
    &\frac{d\hat{E}(t)}{dt}=-P_\mathrm{batt}(t),\;t\in [t_0,t_e],\;\hat{E}(t_0)=E(t_0),\label{eqn:opt_batt_dyn}\\
    &\hat E(t)\in[0,E_\mathrm{max}],\;P_\mathrm{batt}(t)\in[\underline{P}_\mathrm{batt},\overline{P}_\mathrm{batt}],\label{eqn:opt_batt_cons}\\
    &\begin{cases}
    0\leq \overline{u}_i(t)-u_i(t)\leq c_{i}(t),\; &t\in [t_{s_i},t_{s_i}+T_{e_i}]\\[0.12cm]
    u_i(t) = 0,\;&\text{otherwise}
    \end{cases}\label{eqn:opt_input}\\
    & 0\leq t_{s_i}-t_{r_i}\leq \Delta_i^d,\;i \in \I_r(t_0).\label{eqn:opt_cons_delay}\\
    & t_{r_i}= \mathfrak{p}(i),\;i\in \I_r(t_0)\backslash\I_s.\label{eqn:opt_seq_cons}
\end{align}
\end{subequations}
The objective penalizes both the delay and the curtailment, and $\hat{E}(\cdot)$ denotes the virtual counterpart of the actual battery $E(\cdot)$ optimized over the prediction horizon $[t_0,t_e]$. Additionally, the total power consumption constraint~\eqref{eqn:opt_bus_cons} is imposed on $[t_0,t_e]$. Therefore, the end of the horizon $t_e$ is required to be later than the finish time of the all the requests. Note that the start times $t_{s_i}$ are decision variables, we therefore choose $t_e\geq\max_{i\in\I_r(t_0)}\{t_{r_i}+T_{d,i}+T_{e_i}\}$. The constraint~\eqref{eqn:opt_input} ensures that $u_i(t)=0$ when it is not operating. Finally, for those loads outside $\mathcal{I}_s$ constraint~\eqref{eqn:opt_seq_cons} synthetically enforces that a component of an interruptible load can execute only after its preceding load is finished.

\section{Numerical Implementation}\label{sect:num}
Problem~\eqref{eqn:opt_inf} is numerically intractable due to the infinite dimensional decision variable $u_i(\cdot)$. This section discuss\change{es} the transcription of this infinite dimensional problem to a finite dimensional form.
\subsection{Inhomogeneous Time Grids}
The source of the computational intractability of~\eqref{eqn:opt_inf} is the $u_{i}(\cdot)$ in continuous time. Therefore, we adopt the idea of direct optimal control~\cite{bock1984multiple} that relaxes the constraints~\eqref{eqn:opt_bus_cons}-\eqref{eqn:opt_input} to the evaluation on finite time instances. To this end, we introduce the following inhomogeneous time grids:
\subsubsection{Global time grid} we denote it by $\mathcal{T}_g:=\{t_{g,j}\}_{j\in\Z_{0}^{N_g}}$ with 
$t_0=t_{g,0} < t_{g,1}\dots<t_{g,N_g-1}<t_{g,N_g}=t_e$.
This time grid is introduced to evaluate the constraints~\eqref{eqn:opt_bus_cons}, ~\eqref{eqn:opt_batt_dyn}  and~\eqref{eqn:opt_batt_cons} point by point. $\mathcal{T}_g$ shifts with respect to the current time instance $t_0$, and its spacing is user-defined and fixed. 

\subsubsection{Local time grid} we denote it by $\mathcal{T}_i:=\{t_{i,j}\}_{j\in\Z_{0}^{N_{l,i}}}$ with
$t_{s_i}=t_{i,0} < t_{i,1}\dots<t_{i,N_{l,i}-1}<t_{i,N_{l,i}}=t_{s_i}+T_{e_i}$ for the $i$-th load. 
This local grid is used to evaluate the actual input in~\eqref{eqn:opt_input}, and it is defined relative to the starting time $t_{s_i}$ and only the evaluation on these points $\{u_i(t)\}_{t\in\mathcal T_i}$ for all $i\in\mathcal I_r(t_0)$ are optimized. 

The direct benefit from this inhomogeneous time grid is on the implementation side. If we fix the resolution of the local time grid, for example, as $\{j\cdot \delta t_i\}_{j\in\Z_{0}^{N_{l,i}}}$ with $\delta t_i$ the resolution of this time grid, the local time grid is then uniquely defined by $\{t_{s_i}+j\cdot\delta t_i\}_{j\in\Z_{0}^{N_{l,i}}}$ with a single scalar decision variable $t_{s_i}$. Meanwhile, as the time grids are introduced independently to each constraint in~\eqref{eqn:opt_bus_cons} and~\eqref{eqn:opt_input}, the resolution of each time grid can be chosen individually. This can help limit the computational complexity, by reducing the resolution of the time grid on less relevant loads. \change{Finally, an example that visualizes these grids for a numerical example is shown in Figure~\ref{fig:case_study} in Section~\ref{sect:case}.}

\subsection{Input Interpolation}\label{sect:intp}
Note that the local time grid of a running load is fixed in terms of absolute time, while the global time grid is always receding with respect to the current time step. Thus, the local time grids \change{do not} need to overlap the global time grid. In order to evaluate each load on the global time grid, an approximation $\{u_i(t)\}_{t\in\mathcal T_g}$ of the actual input profile $u_i(\cdot)$ is required. This approximation can be numerically given by an interpolation of $\{u_i(t)\}_{t\in\mathcal T_i}$. One classical approach is to use a Lagrange polynomial with order $N_{l,i}$. However, high order polynomials cause high sensitivity with respect to $t_{s_i}$, and thus are not desirable for numerical optimization~\cite{burden2015numerical}. 

In our experiments, we noticed that this high sensitivity issue can lead to convergence failures even if a third order piece-wise polynomial is used. Based on our implementations, there are two interpolation methods that are reliable for the considered problem, piece-wise linear interpolation and kernel interpolation, which is a generalized interpolation scheme that includes polynomial interpolation as a special case~\cite[Chapter 4]{paulsen2016introduction}. The regularity of a kernel regression can be controlled by the choice of its kernel function~\cite[Chapter 2]{paulsen2016introduction}, and one such example of good regularity is the \change{radial basis function} (RBF) kernel $k(x,y) = \exp(-\frac{\lVert x-y \rVert^2}{2\sigma^2})$.
\vspace{-2em}
\begin{remark}\label{rmk:cont}
Only continuous interpolation splines are desirable in solving the load scheduling problem. More specifically, \change{a discontinuous point will zero out the gradient between the global time grid and $t_{s_i}$, leading to a frozen $t_{s_i}$ if a gradient-based solver is used.} As a result, piece-wise constant interpolation is not used in this problem. \change{This is not restrictive as continuous functions can approximate any bounded-variation discontinuous function up to arbitrary accuracy.}
\end{remark}

\vspace{-1.2em}
\subsection{Finite Dimensional Problem}
We interpret the $P_\mathrm{batt}(t)$ by a piece-wise constant input on $\mathcal{T}_g$ such that the solution of~\eqref{eqn:opt_batt_dyn} can be worked out analytically, and we define it by $\hat E^*(t)$ in the following. 
Then, we state our finite order approximated problem by
\vspace{-0.3cm}
\begin{subequations}\label{eqn:opt_finite}
\begin{align}
    \underset{\substack{t_{s_i},\{P_\mathrm{batt}(t)\}_{t\in\mathcal{T}_g}\\\{u_i(t)\}_{i\in\mathcal{I}_r(t_0),t\in\mathcal{T}_i}}}{\mathrm{minimize}} \sum_{i\in\I_r(t)} \Big(J_{i}^d(t_{s_i}-t_{r_i})+\tilde L_{i}^c(u_{i}(\cdot),t_{s_i})\Big)\nonumber
\end{align}
subject to: $\sum_{i\in\I_r(t_0)} u_i(t) \leq P_\mathrm{min}(t)+P_\mathrm{batt}(t),\;t\in\mathcal{T}_g$
\begin{align}
    &\hspace{-3mm}\hat E^*(t)\in[0,E_\mathrm{max}],P_\mathrm{batt}(t_{g,i})\in[\underline{P}_\mathrm{batt},\overline{P}_\mathrm{batt}],\;t\in\mathcal{T}_g,\label{eqn:opt_finite_batt_cons}\\
    &\hspace{-2.5mm}u_i(\cdot) = \mathcal{F}(u_i(t),t\in\mathcal{T}_i),\label{eqn:opt_intp}\\
    &\hspace{-2.5mm}0\leq \overline{u}_i(t)-u_i(t)\leq c_{i},\;i \in \I_r(t_0),\;t\in\mathcal T_i,\\
    &\hspace{-2.5mm}0\leq t_{s_i}-t_{r_i}\leq \Delta_i^d,\;i \in \I_r(t_0),\;t\in\mathcal T_i,\label{eqn:opt_finite_ts}\\
    &\hspace{-2.5mm} t_{r_i}= \mathfrak{p}(i),\;i\in \I_r(t_0)\backslash\I_s,
\end{align}
\end{subequations}
where $\tilde L_{i}^c$ represents the numerical integration of~\eqref{eq::L} which is implemented by the collocation method in the following numerical results~\cite[Chapter 3]{canuto2007spectral}. The non-convexity enters in constraint~\eqref{eqn:opt_intp}, \change{where $\mathcal{F}(\cdot)$ denotes the interpolation of the actual input (Section~\ref{sect:intp})}. The utilization of the absolute time in the numerical implementation is important, as the $t_{r_i}$, $t_{s_i}$ and $t_0$ uniquely defines all the time grids used in this problem. \change{The online operation follows the receding horizon MPC scheme, details can be found in~\cite[Chapter 1.3]{rakovic2018handbook},\cite{lian2020resource}.}
\vspace{-1em}
\begin{remark}
Based on our experiments, it is important to remove redundant decision variables to avoid introducing unnecessary sensitivity to the numerical solver. For example, if a load $i$ is started at $\Tilde{t}$ and has not yet finished, one should avoid adding an extra equality constraint $t_{s_i}=\Tilde{t}$. Instead, the code should take $t_{s_i}$ as a constant $\Tilde{t}$, and discard its related constraints such as~\eqref{eqn:opt_finite_ts}. Based on this idea, the starting time and the past input of a running load are treated as constant, and the code only keeps its future input in the list of decision variables.
\end{remark}
\vspace{-1.2em}

\subsection{Discussions}
To better illustrate the benefit of the continuous formulation, we first review its standard discrete-time mixed integer formulation as follows:
\begin{subequations}\label{eqn:opt_int}
\begin{equation}
\begin{array}{l}    \underset{\substack{t_{s_i},\{P_\mathrm{batt}(t_{m_j})\}\\\{u_i(t_{m_j})\}_{i\in\mathcal{I}_r(t_0)}}}{\mathrm{minimize}}\quad  \sum_{i\in\I_r(t)}\Big(J_{i}^d(t_{s_i}-t_{r_i})\\\label{eqn:opt_int_obj}
\qquad\qquad+\sum_{j=0}^{N_m}T_s\cdot J_{i}^c(\overline{u}_i(t_{m_j})-u_i(t_{m_j}))\Big)
\end{array}
\end{equation}
subject to: $t_{m_j}=t_0+jT_s,\;j\in\mathbb{Z}_0^{N_m}$,
\begin{align}
    &\hspace{-2mm}\sum\limits_{i\in\I_r(t_0)} u_i(t_{m_j}) \leq P_\mathrm{min}(t_{m_j})+P_\mathrm{batt}(t_{m_j}),\\\notag
    &\hspace{-2mm}\hat{E}(t_{m_{j+1}})=\hat{E}(t_{m_j})-T_s\cdot P_\mathrm{batt}(t_{m_j}),\;\hat{E}(t_{m_0})=E(t_0),\\
    &\hspace{-2mm}\hat{E}(t_{m_j})\in[0,E_\mathrm{max}],P_\mathrm{batt}(t_{m_j})\in[\underline{P}_\mathrm{batt},\overline{P}_\mathrm{batt}],\\
    &\hspace{-2mm}\forall\,i \in \I_r(t_0),\;t_{s_i}\in \{t_0+jT_s\}_{j\in\mathbb{Z}},\label{eqn:opt_int_delay}\\
    & \hspace{-2mm} t_{r_i}= \mathfrak{p}(i),\;i\in \I_r(t_0)\backslash\I_s\\
    &\hspace{-2mm} 0\leq t_{s_i}-t_{r_i}\leq \Delta_i^d\label{eqn:opt_int_delay_cons},\\
    &\hspace{-2.6mm}\left\{\begin{aligned}
    &0\leq \overline{u}_i(t_{m_j})-u_i(t_{m_j})\leq c_i,\;t\in[t_{s_i},t_{s_i}+T_{e_i}]\\[0.12cm]
    &u_i(t_{m_j})=0,\text{otherwise}
    \end{aligned}\right.
\end{align}
\end{subequations}
where $N_m$ denotes the prediction horizon and $T_s$ is the sampling time, and the objective~\eqref{eqn:opt_int_obj} approximates the integral of curtailment loss by the Euler method. The decision of the shifting $t_{s_i}$ is determined by an integer variable in~\eqref{eqn:opt_int_delay}. This mixed integer problem~\eqref{eqn:opt_int} was used in~\cite{9147361,di2012model} without considering the interruptible load, and the major difference between the mixed-integer formulation and the proposed continuous problem~\eqref{eqn:opt_finite} lies in the decision of the shifting time $t_{s_i}$ and the complexity of the time grids. In particular, the discrete time formulation only has one grid of the same resolution, i.e., the sampling time $T_s$, among all the loads and constraints. Therefore, the complexity of this shared time grid depends on the highest resolution  that is required for a proper decision of the load profile or the shifting. Accordingly, a finer $T_s$ will significantly increase the numerical complexity of the problem because the number of the integers feasible for constraints~\eqref{eqn:opt_int_delay_cons} increases. In comparison, the proposed scheme in~\eqref{eqn:opt_finite} gets rid of this limitation in two aspects:
\begin{itemize}
    \item The feasible set of shifting $t_{s_i}$ is a continuum and the complexity with respect to the decision of $t_{s_i}$ is fixed regardless of the resolution of any time grids.
    \item Each load has its local time grid, thus any change of its local grid will not affect other time grids.
\end{itemize}

\change{Overall, the continuous formulation is preferable from a computational aspect. It can easily adapt to a problem with nonlinear dynamics, while the mixed integer formulation suffers from the numerical intractability issue as a non-convex mixed integer problem. Furthermore, the choices of solvers with the continuous formulation is more flexible. On one hand, it can use a local algorithm that is easier to deploy in embedded systems with limited computational power~\cite{nocedal2006numerical}.
On the other hand, similar to the mixed integer solvers, it can choose a global solver~\cite{floudas2013deterministic} to ensure global optimality without polynomial time convergence. Last but not least, the continuous formulation is more scalable in that it can use distributed optimization algorithms because the appliances are only coupled via the affine constraint~\eqref{eqn:opt_bus_cons}.}

\section{Numerical Results}\label{sect:result}
We first consider uninterruptible loads to compare the proposed continuous formulation~\eqref{eqn:opt_finite} with the mixed integer version~\eqref{eqn:opt_int}. Then an example with interruptible loads based on real data validates the proposed scheme. We consider the worst case, so the actual generated power is only $P_\mathrm{min}$.
\begin{figure*}[t]
    \centering
    \begin{tikzpicture}
    \begin{groupplot}[
        legend columns=3,
        legend style={
    	font=\footnotesize},
        group style=
            {columns=3,horizontal sep = 3em}]
    \nextgroupplot[xmin=0, xmax=8,
    ymin=-0, ymax= 4.5,
    enlargelimits=false,
    clip=true,
    grid=major,
    mark size=0.5pt,
    width=.36\linewidth,
    height=0.2\linewidth,ylabel = {load 1 $[\mathrm{kW}]$} ,xlabel= {time$[\mathrm{h}]$},
    legend to name=grouplegend,
    label style={font=\scriptsize},
    ticklabel style = {font=\scriptsize},
    ylabel style={at={(axis description cs:-0.05,0.5)}},
    xlabel style={at={(axis description cs:0.5,-0.1)}}]
    \pgfplotstableread{data/compare/compare_continuous1.dat}{\dat}
    \addplot+ [ultra thick, mark=none, mark options={fill=white, scale=1.2},MyBlue] table [x={tr}, y={traj}] {\dat};
    \addlegendentry{Desired profile w.o. delay}

    \addplot+ [ultra thick,dashed, mark=none, mark options={fill=white, scale=1.2},MyRed] table [x={ts}, y={u}] {\dat};
    \addlegendentry{Actual input profile continuous solution}
    
    \pgfplotstableread{data/compare/compare_discrete1.dat}{\dat}
    \addplot+ [ultra thick,dashed, mark=none, mark options={fill=black,draw opacity = 0.7,draw opacity = 0.7, scale=0.5},black,draw opacity = 0.7] table [x={ts}, y={u}] {\dat};
    \addlegendentry{Actual input profile discrete solution}

    \pgfplotstableread{data/compare/compare_continuous3.dat}{\dat}
    \addplot+ [ultra thick, mark=none, mark options={fill=white, scale=1.2},MyBlue] table [x={tr}, y={traj}] {\dat};
    \addplot+ [ultra thick,dashed, mark=none, mark options={fill=white, scale=1.2},MyRed] table [x={ts}, y={u}] {\dat};
    
    \pgfplotstableread{data/compare/compare_discrete3.dat}{\dat}
    \addplot+ [ultra thick,dashed, mark=none, mark options={fill=black,draw opacity = 0.7, scale=0.5},black,draw opacity = 0.7] table [x={ts}, y={u}] {\dat};
    
    \nextgroupplot[xmin=0, xmax=8,
    ymin=-0, ymax= 4.5,
    enlargelimits=false,
    clip=true,
    grid=major,
    mark size=0.5pt,
    width=.36\linewidth,
    height=0.2\linewidth,ylabel = {load 2 $[\mathrm{kW}]$} ,xlabel= {time$[\mathrm{h}]$},
    label style={font=\scriptsize},
    ticklabel style = {font=\scriptsize},
    ylabel style={at={(axis description cs:-0.05,0.5)}},
    xlabel style={at={(axis description cs:0.5,-0.1)}}]
    \pgfplotstableread{data/compare/compare_continuous2.dat}{\dat}
    \addplot+ [ultra thick, mark=none, mark options={fill=white, scale=1.2},MyBlue] table [x={tr}, y={traj}] {\dat};
    \addplot+ [ultra thick,dashed, mark=none, mark options={fill=white, scale=1.2},MyRed] table [x={ts}, y={u}] {\dat};
    
    \pgfplotstableread{data/compare/compare_discrete2.dat}{\dat}
    \addplot+ [ultra thick,dashed, mark=none, mark options={fill=black,draw opacity = 0.7, scale=0.5},black,draw opacity = 0.7] table [x={ts}, y={u}] {\dat};
    
    \pgfplotstableread{data/compare/compare_continuous4.dat}{\dat}
    \addplot+ [ultra thick, solid,mark=none, mark options={fill=white, scale=1.2},MyBlue] table [x={tr}, y={traj}] {\dat};
    \addplot+ [ultra thick, dashed, mark=none, mark options={fill=white, scale=1.2},MyRed] table [x={ts}, y={u}] {\dat};
    
    \pgfplotstableread{data/compare/compare_discrete4.dat}{\dat}
    \addplot+ [ultra thick,dashed, mark=none, mark options={fill=black,draw opacity = 0.7, scale=0.5},black,draw opacity = 0.7] table [x={ts}, y={u}] {\dat};
    
    \pgfplotstableread{data/compare/compare_continuous5.dat}{\dat}
    \addplot+ [ultra thick,solid, mark=none, mark options={fill=white, scale=1.2},MyBlue] table [x={tr}, y={traj}] {\dat};
    \addplot+ [ultra thick,dashed, mark=none, mark options={fill=white, scale=1.2},MyRed] table [x={ts}, y={u}] {\dat};
    
    \pgfplotstableread{data/compare/compare_discrete5.dat}{\dat}
    \addplot+ [ultra thick,dashed, mark=none, mark options={fill=black,draw opacity = 0.7, scale=0.5},black,draw opacity = 0.7] table [x={ts}, y={u}] {\dat};
    
    \nextgroupplot[xmin=0, xmax=8,
    ymin=-0, ymax= 0.65,
    enlargelimits=false,
    clip=true,
    grid=major,
    mark size=0.5pt,
    width=.36\linewidth,
    height=0.2\linewidth,ylabel = {battery level $[\mathrm{kWh}]$},xlabel= time$\begin{bmatrix}\text{h}\end{bmatrix}$,
    label style={font=\scriptsize},
    ticklabel style = {font=\scriptsize},
    ylabel style={at={(axis description cs:-0.1,0.5)}},
    xlabel style={at={(axis description cs:0.5,-0.1)}}]
    
    \pgfplotstableread{data/compare/compare_continuous_batt.dat}{\dat};
    \addplot+ [ultra thick, dashed, mark=none, mark options={fill=white, scale=1.2},MyRed] table [x={t}, y={e}] {\dat};
    
    \pgfplotstableread{data/compare/compare_discrete_batt.dat}{\dat};
    \addplot+ [ultra thick, dashed, mark=none, mark options={fill=white, scale=1.2},black,draw opacity = 0.7] table [x={t}, y={e}] {\dat};
    
    \end{groupplot}
    \path (group c1r1.north east) -- node[above]{\ref{grouplegend}} (group c3r1.north west);
    \end{tikzpicture}
   
    \caption{Comparison of closed loop behaviour between continuous-time~\eqref{eqn:opt_finite} and mixed-integer formulations~\eqref{eqn:opt_int} with $T_s = 0.1[\text{h}]$}
    \label{fig:compare}
    \vspace{-2em}
\end{figure*}
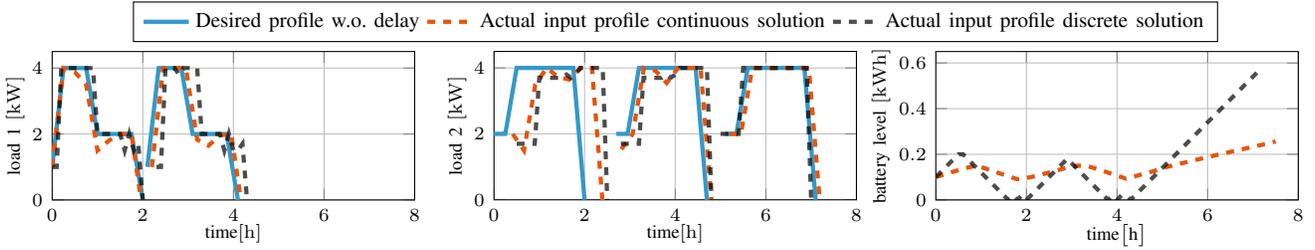
\subsection{Benchmark Comparison}
This example considers uninterruptible loads, and the proposed scheme is compared against the benchmark mixed-integer formulation~\eqref{eqn:opt_int}, which is implemented with \texttt{Yalmip} interfacing the \texttt{Gurobi} MIQP solver~\cite{gurobi}, one of the fastest commercial solvers, and \textsc{BNB}, an open source mixed integer solver. Moreover, the continuous problem~\eqref{eqn:opt_finite} is implemented with \texttt{Casadi}~\cite{andersson2019casadi} interfacing the \texttt{Ipopt} solver~\cite{wachter2006implementation}. All code is run on a laptop with Intel i7-11800H and 32 GB of memory.
We optimize the implementation of the mixed integer formulation by an auxiliary system~\cite{9147361}\footnote{\change{This pure delayed system outputs the load profile once started, which converts an integer delay variable to a sequence of binary decision variables.}}, without which solution time is reported to be significantly longer~\cite{9147361}.

In this comparison, two loads are considered, whose load profiles are plotted as blue solid lines in Figure~\ref{fig:compare}. The shifting time upper bound and the curtailment upper bound are, $\Delta_i^d=1\,[\mathrm{h}]$ and $c_i=0.5\change{[\mathrm{kW}]}$. The shifting loss and the curtailment loss are $J_{i}^d = 10\max\{t_{s_i}-t_{r_i}-0.1,0\}^2,\;J_{i}^c = \left(\overline{u}_i(t)-u_i(t)\right)^2$,
where the shifting loss is zero if the shifting time is less than $0.1\,[\mathrm{h}]$. Accordingly, to ensure that the discrete version~\eqref{eqn:opt_int} can discover this property, its sampling time $T_s$ is set below $0. 1\,[\mathrm{h}]$. To make good use of the inhomogeneous time grid system, the local time grid in the continuous formulation~\eqref{eqn:opt_finite} has a resolution of $0.2\,[\mathrm{h}]$, with a global time grid of resolution $0.1\,[\mathrm{h}]$. The load of each appliance is interpolated by kernel regression with the RBF kernel. The total power consumption is upper bounded by $P_\mathrm{min}(\cdot)=5.5\change{[\mathrm{kW}]}$. Meanwhile, the battery is bounded by $\underline{P}_\mathrm{batt}=-0.2\change{[\mathrm{kW}]}$, $\overline{P}_\mathrm{batt}=0.2\change{[\mathrm{kW}]}$, $E_\mathrm{max}=1\change{[\mathrm{kWh}]}$.

Regarding the aforementioned $P_\mathrm{min}(\cdot)$ and the battery dynamics, if these two loads are requested to start at the same time, the only feasible solution is to shift the second load and curtail at least one of them. A sequence of requests are sent to the system, and the closed-loop simulation 
\change{results} are plotted Figure~\ref{fig:compare}. It is observed that both schemes successfully shift the second load and curtail one of them to ensure feasibility. \change{However}, the continuous scheme tends to introduce more curtailment \change{which} results in less delay. The statistics of the proposed continuous formulation and the mixed integer formulation with different sampling time\change{s} are summarized in Table~\ref{tab:sol_time}. In particular, both
the continuous and the discrete version are initialized by a
random feasible shifting, and the solution time is calculated
by the average of 50 independent runs. As the mixed-integer formulation solves an MIQP and can therefore find a global optimum, we can see that the continuous formulation results in a comparable performance, while using significantly less solution time. Meanwhile, we can see that the solution time of the mixed-integer formulation is sensitive to the sampling time and solver, even though smaller sampling time can improve its optimality.
\begin{table}[htbp!]
    \centering
    \footnotesize
    \begin{tabular}{ccccc}
    \hline
    &\shortstack{Solution Time\\(Gurobi/BNB)}& \shortstack{Delay \\Loss}&\shortstack{Curtailment\\Loss}&\shortstack{Total\\Loss}\\   \hline
    Continuous&  1.1011\,[s]& 1.2181 & 0.2525&1.4706\\\hline
    \shortstack{MIQP\\($T_s=0.1[\text{h}]$)}&  \shortstack{9.1340\,[s]\\49.2573\;[s]} & 1.300 & 0.1440&1.4440\\\hline
    \shortstack{MIQP\\($T_s=0.05[\text{h}]$)}& \shortstack{16.6564\,[s]\\194.34\;[s]}& 1.275  & 0.1450&1.4200\\\hline
    \shortstack{MIQP\\($T_s=0.025[\text{h}]$)}&  \shortstack{36.9109\,[s]\\N.A} & 1.2875  & 0.1215& 1.4090\\\hline
    \end{tabular}
    \caption{Statistics of two formulations: loss is the accumulated actual loss evaluated on all the closed-loop input profiles. The solution time for BNB with $T_s = 0.025[\text{h}]$ is dropped as each run takes more than 2 hours.}
    \label{tab:sol_time}
\end{table}

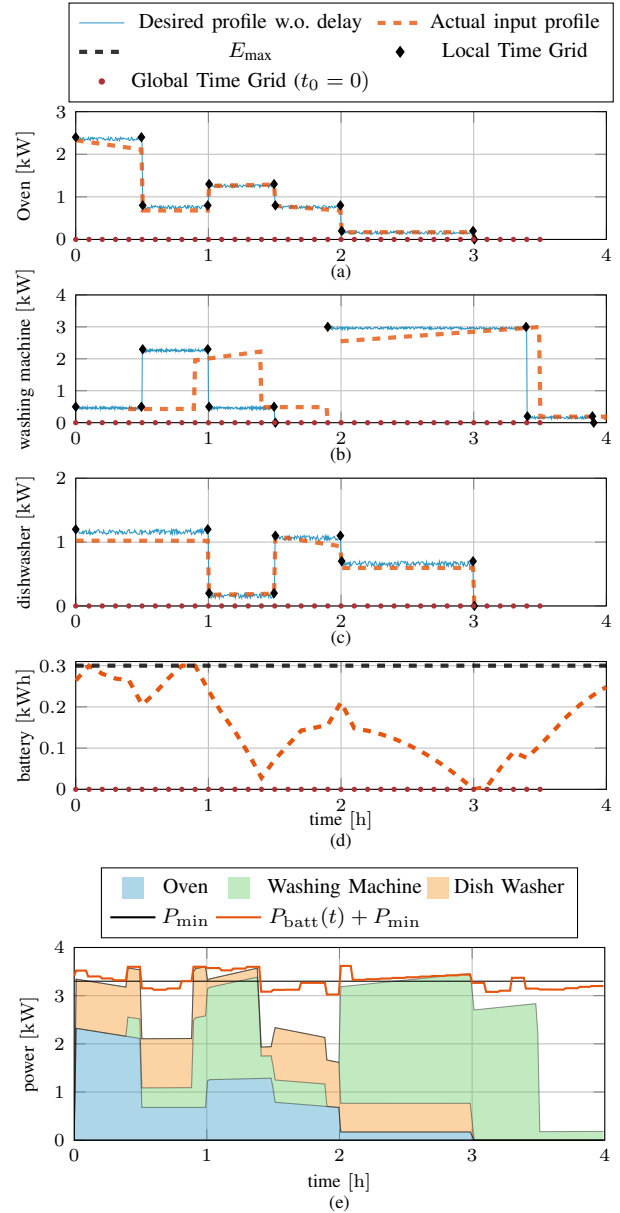
\begin{figure}[htbp!]
    \centering
    \begin{tikzpicture}
    \begin{groupplot}[
        legend columns=2,
        legend style={
    	font=\footnotesize},
        group style=
            {group size= 1 by 4,
            vertical sep=2.1em}]
    \nextgroupplot[xmin=0, xmax=4,
    ymin=-0, ymax= 3,
    enlargelimits=false,
    clip=true,
    grid=major,
    mark size=0.5pt,
    width=\linewidth,
    height=0.38\linewidth,ylabel = {Oven $[\mathrm{kW}]$},
    xlabel style={align=center},xlabel= (a),
    legend to name=grouplegend2,
    label style={font=\scriptsize},
    ticklabel style = {font=\scriptsize},
    ylabel style={at={(axis description cs:-0.06,0.5)}},
    xlabel style={at={(axis description cs:0.5,-0.12)}}]
    \pgfplotstableread{data/case/load1.dat}{\dat}
    \addplot+ [mark=none, mark options={fill=white, scale=1.2},MyBlue] table [x={t}, y={u}] {\dat};
    \addlegendentry{Desired profile w.o. delay}
    
    \pgfplotstableread{data/case/case1.dat}{\dat}
    \addplot+ [ultra thick,dashed,draw opacity = .8, mark=none, mark options={fill=white, scale=1.2},MyRed] table [x={ts}, y={u}] {\dat};
    \addlegendentry{Actual input profile}
    
    \addlegendimage{line legend,ultra thick,dashed,draw opacity=0.8,mark=none, mark options={fill=white, scale=1.2},black}
    \addlegendentry{$E_{\text{max}}$}  
    
    \pgfplotstableread{data/case/grid_local1.dat}{\dat}
    \addplot+ [only marks, mark=diamond*, mark options={fill=black, scale=4,draw opacity=0},forget plot] table [x={t}, y={y}] {\dat};
    
    \addlegendimage{only marks, mark=diamond*,color=black,mark options={scale=1}}
    \addlegendentry{Local Time Grid}
    
      \pgfplotstableread{data/case/grid_global.dat}{\dat}
    \addplot+ [only marks, mark=*, mark options={fill=Maroon, scale=2,draw opacity=0},forget plot] table [x={t}, y={y}] {\dat};
    \addlegendimage{only marks, mark=*,color=Maroon,mark options={scale=0.5}}
    \addlegendentry{Global Time Grid ($t_0=0)$}

    \nextgroupplot[xmin=0, xmax=4,
    ymin=-0, ymax= 4,
    enlargelimits=false,
    clip=true,
    grid=major,
    mark size=0.5pt,
    width=\linewidth,
    height=0.38\linewidth,ylabel = {washing machine $[\mathrm{kW}]$},
    xlabel style={align=center},xlabel= (b),
    label style={font=\scriptsize},
    ticklabel style = {font=\scriptsize},
    ylabel style={at={(axis description cs:-0.06,0.5)}},
    xlabel style={at={(axis description cs:0.5,-0.12)}}]
    \pgfplotstableread{data/case/load2.dat}{\dat}
    \addplot+ [mark=none, mark options={fill=white, scale=1.2},MyBlue] table [x={t}, y={u}] {\dat};
    
    \pgfplotstableread{data/case/case2.dat}{\dat}
    \addplot+ [ultra thick,dashed,draw opacity = .8, mark=none, mark options={fill=white, scale=1.2},MyRed] table [x={ts}, y={u}] {\dat};
    
    \pgfplotstableread{data/case/grid_local2.dat}{\dat}
    \addplot+ [only marks, mark=diamond*, mark options={fill=black, scale=4,draw opacity=0}] table [x={t}, y={y}] {\dat};
    
    \pgfplotstableread{data/case/grid_global.dat}{\dat}
    \addplot+ [only marks, mark=*, mark options={fill=Maroon, scale=2,draw opacity=0},forget plot] table [x={t}, y={y}] {\dat};

    \pgfplotstableread{data/case/load4.dat}{\dat}
    \addplot+ [mark=none, mark options={fill=white, scale=1.2},MyBlue] table [x={t}, y={u}] {\dat};
    
    \pgfplotstableread{data/case/case4.dat}{\dat}
    \addplot+ [ultra thick,dashed,draw opacity = .8, mark=none, mark options={fill=white, scale=1.2},MyRed] table [x={ts}, y={u}] {\dat};
    
    \pgfplotstableread{data/case/grid_local4.dat}{\dat}
    \addplot+ [only marks, mark=diamond*, mark options={fill=black, scale=4,draw opacity=0},forget plot] table [x={t}, y={y}] {\dat};
    
    \nextgroupplot[xmin=0, xmax=4,
    ymin=-0, ymax= 2,
    enlargelimits=false,
    clip=true,
    grid=major,
    mark size=0.5pt,
    width=\linewidth,
    height=0.38\linewidth,ylabel = {dishwasher $[\mathrm{kW}]$},
    ytick distance = 1,
    xlabel style={align=center},xlabel= (c),
    label style={font=\scriptsize},
    ticklabel style = {font=\scriptsize},
    ylabel style={at={(axis description cs:-0.06,0.5)}},
    xlabel style={at={(axis description cs:0.5,-0.12)}}]
    \pgfplotstableread{data/case/load3.dat}{\dat}
    \addplot+ [mark=none, mark options={fill=white, scale=1.2},MyBlue] table [x={t}, y={u}] {\dat};
    
    \pgfplotstableread{data/case/case3.dat}{\dat}
    \addplot+ [ultra thick,dashed,draw opacity = .8, mark=none, mark options={fill=white, scale=1.2},MyRed] table [x={ts}, y={u}] {\dat};
    
    \pgfplotstableread{data/case/grid_local3.dat}{\dat}
    \addplot+ [only marks, mark=diamond*, mark options={fill=black, scale=4,draw opacity=0},forget plot] table [x={t}, y={y}] {\dat};
    
    \pgfplotstableread{data/case/grid_global.dat}{\dat}
    \addplot+ [only marks, mark=*, mark options={fill=Maroon, scale=2,draw opacity=0},forget plot] table [x={t}, y={y}] {\dat};
    
    \nextgroupplot[xmin=0, xmax=4,
    ymin=-0, ymax= 0.31,
    enlargelimits=false,
    clip=true,
    grid=major,
    mark size=0.5pt,
    width=\linewidth,
    height=0.38\linewidth,ylabel = {battery $[\mathrm{kWh}]$},
     ytick distance = .1,
    xlabel style={align=center},xlabel= {time $[\mathrm{h}]$}\\(d),
    label style={font=\scriptsize},
    ticklabel style = {font=\scriptsize},
    ylabel style={at={(axis description cs:-0.06,0.5)}},
    xlabel style={at={(axis description cs:0.5,-0.1)}}]
    \pgfplotstableread{data/case/case_batt.dat}{\dat}
    \addplot+ [ultra thick,dashed,mark=none, mark options={fill=white, scale=1.2},MyRed] table [x={t}, y expr=\thisrow{e}] {\dat};
    \addplot+ [ultra thick,dashed,draw opacity=0.8,mark=none, mark options={fill=white, scale=1.2},black] table [x={t}, y expr=\thisrow{e_max}] {\dat};
    
    \pgfplotstableread{data/case/grid_global.dat}{\dat}
    \addplot+ [only marks, mark=*, mark options={fill=Maroon, scale=2,draw opacity=0},forget plot] table [x={t}, y ={y}] {\dat};
    \end{groupplot}
    
    \path (group c1r1.north east) -- node[above]{\ref{grouplegend2}} (group c1r1.north west);
    
    \end{tikzpicture}\\
    \begin{tikzpicture}
        \begin{groupplot}[
        legend columns=3,
        legend style={
    	font=\footnotesize},
        group style=
            {columns=1,}]

    \nextgroupplot[xmin=0, xmax=4,
    ymin=-0, ymax= 4,
    enlargelimits=false,
    clip=true,
    grid=major,
    mark size=0.5pt,
    legend to name=grouplegend2,
    width=1\linewidth,
    height=0.48\linewidth,
    ylabel style={align=center},
    ylabel = {power $[\mathrm{kW}]$},
    xlabel style={align=center},
    xlabel= {time $[\mathrm{h}]$}\\(e),
    label style={font=\scriptsize},
    ticklabel style = {font=\scriptsize},
    ylabel style={at={(axis description cs:-0.05,0.5)}},
    xlabel style={at={(axis description cs:0.5,-0.12)}}]
    
    \pgfplotstableread{data/case/case_chart.dat}{\dat}
    \addplot+ [name path=A,mark=none, mark options={fill=white, scale=1.2},black,forget plot] table [x={t}, y={sum1}] {\dat};
    
    \addplot+ [name path=B,mark=none, mark options={fill=white, scale=1.2},black, draw opacity = .4,forget plot] table [x={t}, y={sum2}] {\dat}; 
    \addplot+ [name path=C,mark=none, mark options={fill=white, scale=1.2},black, draw opacity = .4,forget plot] table [x={t}, y={sum3}] {\dat}; 
    \addplot+ [name path=D,mark=none, mark options={fill=white, scale=1.2},black, draw opacity = .4,forget plot] table [x={t}, y={sum4}] {\dat}; 
    \addplot+ [name path=E,mark=none, mark options={fill=white, scale=1.2},black, draw opacity = .4,forget plot] table [x={t}, y={sum5}] {\dat}; 
    \addplot[MyBlue,fill opacity=0.4,forget plot] fill between[of=A and B]; 
    \addplot[MyGreen,fill opacity=0.4,forget plot] fill between[of=B and C]; 
    \addplot[BurntOrange,fill opacity=0.4,forget plot] fill between[of=C and D]; 
    \addplot[MyGreen,fill opacity=0.4,forget plot] fill between[of=D and E];

    \addplot+ [thin, mark=none, mark options={scale=1},black,forget plot] table [x={t}, y={u_max}] {\dat};
    
    \addplot+ [thick, mark=none, mark options={fill=white, scale=1.2},MyRed,forget plot] table [x={t}, y={p}] {\dat}; 
     
     \addlegendimage{only marks, mark=square*,color=MyBlue,mark options={scale=2,draw opacity=0},fill opacity = .4}
    \addlegendentry{Oven}
    \addlegendimage{only marks, mark=square*,color=MyGreen,mark options={scale=2,draw opacity=0},fill opacity = .4}
    \addlegendentry{Washing Machine}
    \addlegendimage{only marks, mark=square*,color=BurntOrange,mark options={scale=2,draw opacity=0},fill opacity = .4}
    \addlegendentry{Dish Washer}
    \addlegendimage{line legend,thick, solid, mark=none, black}
    \addlegendentry{$P_{\mathrm{min}}$}
    \addlegendimage{line legend,thick, mark=none, mark options={fill=white, scale=1}, MyRed}
    \addlegendentry{$P_{\mathrm{batt}}(t)+P_{\mathrm{min}}$}
            
    \end{groupplot}
    \path (group c1r1.north east) -- node[above]{\ref{grouplegend2}} (group c1r1.north west);
    \end{tikzpicture}

    \pgfplotstableread{data/compare/compare_continuous3.dat}{\dat}
    \caption{\change{Case study of home appliance scheduling. (a)-(c): Individual appliance profiles with local and global time grid markers. (d) Battery Energy. The global time grid is plotted on the horizontal axis in (a)-(d) to show the correspondence between the local and global grids and the load profile. The global grid is shown as used at $t_0=0$; the grid is shifts with respect to $t_0$.} (e): Stacked chart of the power consumption of all the appliances. The curve of $P_{\mathrm{batt}}(t)+P_{\mathrm{min}}$ shows the change of the power supply profile and how the power limit constraint~\eqref{eqn:opt_bus_cons} is satisfied. The input trajectories are smooth as they are the planned input profiles.}
    \label{fig:case_study}
\end{figure}
\subsection{Case Study with Interruptible Loads}\label{sect:case}


We now consider the scheduling of three home appliances: an oven, washing machine and dishwasher, within which the washing machine is interruptible between the washing and the drying program. The load profiles based on real-world data are plotted as thick blue lines Figure~\ref{fig:case_study} (a)-(c)~\cite{data}, where the washing program of the washing machine takes 1.5 hours and the drying program takes another 2 hours. To reduce the model complexity, we filter out the fluctuations and use a piece-wise constant function to model the desired load profile. These load profiles have significantly different patterns so we use the proposed inhomogeneous time grid system to minimize the number of decision variables. The global grid is equidistant with a resolution of $0.1\,[\mathrm{h}]$, while each load has a non-equidistant local time grid chosen to capture the major patterns in the desired profile, shown with diamonds in Figure~\ref{fig:case_study}. The time grids are: 
\begin{align*}
    &\text{Oven:}\;[0,0.495,0.5, 0.995,1.0,1.495,1.5,1.995,2,2.99,3]\\
    &\substack{\text{Washing}\\\text{Machine}}:  
    \begin{cases}
     [0,0.495,0.5,0.995,1.0,1.495,1.5]\\
     [0,1.495,1.5,1.995,2]
    \end{cases}\\
    &\text{Dishwasher:}\;[0,0.995,1,1.495,\change{1.5},1.995,2,2.995,3]\;.
\end{align*}
A piece-wise linear interpolation is used. The $P_\mathrm{min}$ is $3.2\change{[\mathrm{kW}]}$ with a battery dynamics confined by $\underline{P}_\mathrm{batt}=-0.2\change{[\mathrm{kW}]}$, $\overline{P}_\mathrm{batt}=0.2\change{[\mathrm{kW}]}$, $E_\mathrm{max}=0.3\change{[\mathrm{kWh}]}$. Meanwhile, the delay penalty and the curtailment penalty are defined by $J_{d_i} = 10(t_{s_i}-t_{r_i})^2,\; J_{c_i} = 0.1\left|\overline{u}_i(t)-u_i(t)\right|$.

The result of this scheduling problem is shown in Figure~\ref{fig:case_study}, from which it is observed that the oven is launched without any delay, and the washing machine is shifted to ensure the satisfaction of the power limit. It is also observed the battery is pre-charged at around $0.5\,[\mathrm h]$ and $1.4\,[\mathrm h]$ and then discharged at around $0.9\,[\mathrm h]$ and $2\,[\mathrm h]$ to compensate the peak load in the washing and the drying program. It is noteworthy to point out the curtailment adapts to the available power. This can be observed from the curtailment of the drying program, which is only introduced at around $2\,[\mathrm h]$ but not $3.5\,[\mathrm h]$. \change{This} is because the dishwasher finished at $3\,[\mathrm h]$ and the power supply can sustain the drying program. 
\vspace{-0.1em}
\section{Conclusions}\label{sect:conclusion}
This work proposes a continuous-time formulation of a load scheduling problem within a household. An inhomogeneous time grid system is introduced to decouple the model complexity of each load. The advantage over the standard mixed integer version is validated by a numerical example with better solution time and \change{ comparable closed-loop performance}. Finally, the proposed scheme is validated in a scheduling problem consisting of an oven, an interruptible washing machine and a dishwasher.
\vspace{-0.12cm}

\tiny{
\bibliographystyle{ieeetr}
\bibliography{ref.bib}}

\end{document}